\documentclass{aa}
\usepackage{epsfig}

\newcommand{\lesssim}{\mathrel{\hbox{\rlap{\hbox{\lower4pt\hbox{$\sim$}}}\hbox{$<$}}}}
\newcommand{\gesssim}{\mathrel{\hbox{\rlap{\hbox{\lower4pt\hbox{$\sim$}}}\hbox{$>$}}}}

\newcommand{\bmv}{B$-$V}

\newcommand{\teff}{T$_{\rm eff}$}
\newcommand{\nli}{$\log$~n(Li)}

\begin{document}

\title{The evolution of lithium depletion in young open clusters:
NGC~6475
\thanks{Based on observations collected at the European Southern Observatory}}

   \subtitle{ }

   \author{P. Sestito\inst{1} \and S. Randich\inst{2} \and 
        J.-C. Mermilliod\inst{3} 
         \and R. Pallavicini\inst{4} }

   \offprints{P. Sestito, email:sestito@arcetri.astro.it}

\institute{Dipartimento di Astronomia, 
Universit\`a di Firenze, Largo E. Fermi 5,
            I-50125 Firenze, Italy
\and
INAF/Osservatorio Astrofisico di Arcetri, Largo E. Fermi 5,
             I-50125 Firenze, Italy
\and
Institut d'~Astronomie, Universit\'e de Lausanne, CH-1290,
Chavannes--des--Bois, Switzerland
\and
INAF/Osservatorio Astronomico di Palermo, Piazza del
                  Parlamento 1, I-90134 Palermo, Italy}

\titlerunning{Li in NGC~6475}
\date{Received Date: Accepted Date}

\abstract{We have carried out a high resolution spectroscopic survey
of the 220--250 Myr old cluster \object{NGC~6475}: our main
purpose is to investigate Li evolution during the early stages
of the Main Sequence. We have determined Li abundances for 33 late
F to K--type X--ray selected cluster candidates, extending the samples already available in the 
literature; for part of the stars we obtained radial and rotational
velocities, allowing us to confirm the membership and to check
for binarity. We also estimated the cluster metallicity
which turned out to be over--solar ([Fe/H]$\rm{=+0.14 \pm 0.06}$).
Our Li analysis evidenced that (i) late F--type stars 
(\teff$\gesssim$6000~K) undergo a
very small amount of Li depletion during the early phases on the ZAMS;
(ii) G--type stars (6000$\gesssim$\teff$\gesssim$5500~K)
instead do deplete lithium soon after arrival on the ZAMS.
Whereas this result is not new, we show that the time scale for Li
depletion in these stars is almost constant
between 100 and 600 Myr;
(iii) we confirm that the spread observed in early K--type stars
in younger clusters has converged by 220 Myr. No constraints can be
put on later--type stars.
(iv) Finally, we investigate the effect of metallicity on Li depletion
by comparing NGC~6475 with the similar age cluster M~34, but we show that
the issue remains open, given the uncertain metallicity of the latter
cluster. By using the combined NGC~6475+M~34 sample together with the
Hyades and the Pleiades, we compare quantitatively Li evolution from
the ZAMS to 600 Myr with theoretical predictions of standard models.
}
\maketitle
\section{Introduction}\label{intro}
Studies of lithium abundances in stars are very important since Li survives
only in the outermost layers of a star, due
to its low burning temperature:
for this reason, Li
is a good tracer of mixing mechanisms occurring in stellar interiors
during the various phases of stellar evolution.
Several Li surveys in open clusters have been carried out
during the last two decades, in order to investigate Li 
destruction processes and mixing mechanisms, and their
dependence on age, mass and chemical composition.
The resulting empirical picture has evidenced several features
that cannot be explained with standard theoretical models
that include only convection as a mixing mechanism 
(see e.g. the review by Pasquini \cite{pas00} and references therein).

Standard models predict a certain amount of depletion
during Pre--Main Sequence (PMS) evolution of solar--type stars
(e.g. D'~Antona \& Mazzitelli
\cite{dm94}) 
and no depletion at all after arrival on the Zero--Age Main Sequence
(ZAMS). The predicted PMS Li depletion increases as mass decreases
while the predicted Main Sequence (MS) depletion remains small in all cases,
apart from the coolest stars;
moreover Li depletion should depend only on age, chemical composition
and mass (or effective temperature), i.e. stars with the same mass
in a given cluster should all have the same Li abundance.
As far as the Li--metallicity relationship is concerned,
standard models predict that increased metal abundances should lead
to a significant increase of Li depletion during PMS contraction for stars
cooler than $\sim$6000 K (e.g. Chaboyer et al.~\cite{CDP95},
Swenson et al.~\cite{swenson94}): this is due to the fact that
the gas opacity grows up in stars with a higher iron content, thus
the depth of the convective zone (CZ) increases leading to a large
amount of Li depletion. It is worth of mention that also
oxygen, as well as other $\rm{\alpha}$ elements,
largely contributes to
affect the opacity values
and thus the depth of the CZ (Piau \& Turck-Chi\`eze
\cite{pt02}).

The predictions 
of standard models are in contrast with observational results.
Namely,
focusing on the evolution of Li up to the \object{Hyades} 
age ($\sim$ 600 Myr),
observations of very young clusters (ages 30--50 Myr)
show that solar analogs
undergo very little (if any) Li depletion during the 
PMS (e.g. Mart\'\i n \&
Montes \cite{mm97}; Randich et al.~\cite{R97});
on the other hand, the comparison of clusters
of different ages clearly shows 
that these stars do deplete Li while on the MS.
Furthermore,
the star--to--star scatter in Li abundance seen in
young cluster
for stars cooler than $\sim$ 5500~K 
(e.g. Soderblom et al.~\cite{soder93}--hereafter 
S93; Garc\'\i a L\'opez et al.~\cite{gar94}; Jones et al.~\cite{jon96}; 
Randich et al.~\cite{R98}) is clearly in contrast with standard model
predictions;
this dispersion is already present at arrival on the
ZAMS and has disappeared by the age of 
the Hyades ($\sim$ 600--700 Myr, Thorburn et al.~\cite{thor93}).
Based on the observed Li--rotation relationship (S93), and specifically on 
the fact that rapid rotators in young clusters have 
on average higher 
Li abundances than slow rotators, the most commonly accepted 
explanation for the scatter is that Li depletion is
connected to rotationally driven mixing and angular momentum transport.
Note however that slow rotators in young clusters can 
have either high or low Li abundances, as showed by
Randich et al. (\cite{R98}) for \object{$\alpha$ Persei},
and that the Li--rotation relationship
breaks down for the coolest stars (\teff$\lesssim$4500~K, 
Garc\'\i a L\'opez et al.~\cite{gar94}).

Finally, the fact that
no Li depletion--metallicity relationship has so far been convincingly 
demonstrated
(see for example Jeffries \& James \cite{blanco1}), 
at least for~\teff$\gesssim$4700~K, appears in contrast with model
predictions.

In summary, the question remains which mechanism(s) drives or inhibits
Li depletion in stars of different masses during the PMS and MS
phases.

In order to put additional empirical constraints
on early--MS Li depletion processes, it is necessary to enlarge the 
database of  Li observations in young clusters;
for this reason, we carried out a survey of NGC~6475, 
a well populated Southern hemisphere cluster, 
with reported over--solar metallicity.
NGC~6475 is a very good target, since its age of $\sim$ 220 Myr
(Meynet et al.~\cite{mey93}) is
intermediate between those of the \object{Pleiades} and of the Hyades
and it is the closest and most compact open cluster at that age
(distance $\sim$ 250 pc); the estimated spectroscopic iron abundance is
$\mathrm{[Fe/H]=+0.11\pm0.034}$ and the reddening E(B$-$V)=0.06 (James \& Jeffries~\cite{JJ97}--hereafter JJ97).
Li data for this cluster allow us to investigate 
early--MS Li depletion 
and its time scale for solar--type
and lower mass stars, as well as, to some extent,
%2) the presence of a scatter in Li for stars
%with the same \teff, and its evolution with age by comparing data for clusters of
%different ages;
the
dependence of Li depletion on metallicity by comparing NGC~6475 to
\object{M~34} (NGC~1039), surveyed by Jones et al. (\cite{jones97}). 
The latter cluster is about co--eval to NGC~6475 ($\sim$ 250 Myr, 
Jones \& Prosser \cite{jp96});  Canterna et al.~(\cite{canterna})
found a solar metallicity for M~34, based on multicolor ubvy photometry
of two F--type stars, while a recent and more detailed 
high resolution spectroscopic analysis
by Schuler et al.~(\cite{schuler})
evidenced that the iron content could be 
over--solar
([Fe/H]$\rm{=+0.07 \pm 0.04}$). Note that the result
of Schuler et al.~(\cite{schuler}) is based on five solar--type stars;
had they considered their whole sample of nine stars with
$\rm{{4750}\leq{T_{eff}}\leq{6130}}$ K, they would have found 
[Fe/H]$\rm{=+0.02 \pm 0.02}$ (see the quoted reference for more details).

Previous studies of Li in NGC~6475 were carried out by
James \& Jeffries~\cite{JJ97} and James et al.~\cite{J00}
(hereafter J00): the first one
is based on an X--ray selected sample, while in the latter one an
optically selected sample is studied. In this paper, we present the results of
additional Li observations of NGC~6475: our data, merged with the ones
of the two previous works, provide a larger and more statistically significant
sample of stars to further address the issue of Li evolution between the age of
Pleiades and that of the Hyades. In addition, our sample contains a few more 
stars of later spectral--types than the previous surveys, allowing us to 
get some 
insights on early--MS Li depletion for the coolest stars.

In Sect.~2 we describe our sample and the observations; in Sect.~3 
we summarize the radial velocities and abundances analysis, while
the results and a discussion are presented in Sects.~4 and 5.
Finally our conclusions (Sect.~6) close the paper.
\section{Observations}
\begin{figure*}
\psfig{figure=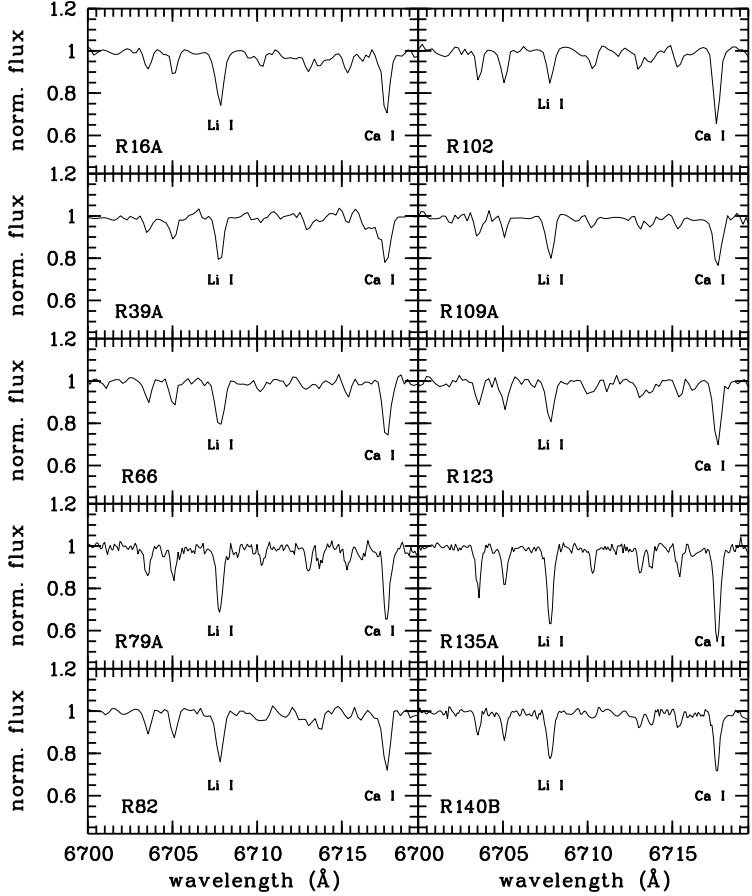, width=14.0cm, angle=-90}
\caption{Sample spectra in the Li region.}\label{spectra}
\end{figure*}

Our original sample includes 34 cluster candidates with 
0.50~$\mathrm{\leq{(B-V)_{0}}\leq}$~1.44, 
selected from the X--ray survey of Prosser
et al.~\cite{pros95} (hereafter P95).
Target stars and photometry are listed in the first three columns
of Table~\ref{original}: the identification number
(Col.~1) and the photometry (Cols.~2 and~3) were retrieved from P95;
a reddening E(B$-$V)=0.06 was adopted.

The observations were carried out during three runs (April 1994, April 1995 and July 1996) at the European Southern Observatory (ESO), La Silla, Chile, 
with the 3.6 m telescope
equipped with CASPEC. During the April 1994 and April 1995 
observing runs
the standard echelle grating
(31.6 lines $\rm{{mm}^{-1}}$) with the red cross--disperser
(158 lines $\rm{mm^{-1}}$) and the short camera 
%(focal length$=$291 mm, f/1.46) 
were used, together with ESO CCD \#32 (TK512, with 512 $\times$ 512
pixels of 27 $\rm {\mu}$);
the nominal resolving power was R$\sim$20,000 (slit aperture of 280 
$\rm {\mu{m}}$). In the July 1996 run the long camera
%(focal length$=$.. {\bf check}) 
and the ESO CCD \#37 (TK1024, with 1024 $\times$ 1024
pixels of 24 $\rm {\mu}$) were used
with a slit aperture of 200 $\rm {\mu{m}}$,
which provided a slightly larger resolving power, R$\sim$29,000.
Exposure times ranged between 10 minutes and 1.5 hours
resulting in S/N ratios of 50--150. For each star, the corresponding
CASPEC run and exposure time are listed in Cols.~4 and 5
of Table~\ref{original}.

About 80\% of the stars were also observed with 
the CORAVEL instrument 
(Baranne et al.~\cite{bmp}) at the 1.54 m Danish telescope at ESO.
Three observations were obtained through the period 1985--1995 
for stars belonging to a program designed to search for faint members,
and one or two measurements were obtained in April and July 1996 for stars 
selected from the ROSAT X--ray source catalog (P95) during the course of 
a long--term systematic program on cluster red dwarfs started in Chile in 
1983. We mention that part of the stars of the JJ97 sample were also
observed by us with CORAVEL.
The CORAVEL observations allowed us to derive
radial velocities and to check for binarity; for part of the 
stars the projected rotational velocities (V~$\sin$~i)
were also determined.
%since we were able to obtain the v $\sin$ i only for a small number of stars, we cannot use these values in our discussion.

CASPEC data reduction was performed with the package MIDAS in 
the ECHELLE context,
following the usual steps: bias subtraction, order definition, order extraction, inter--order background subtraction, flat--fielding and wavelength calibration.
Figure~\ref{spectra} shows examples of normalized spectra 
in the Li region.
\begin{table*}[!ht] \footnotesize
\caption{Photometry, exposure times, radial and rotational velocities 
and membership for target stars selected from P95.}\label{original}
\begin{tabular}{ccccccccc}
\hline
\hline
no. & V & $\mathrm{(B-V)_{0}}$ &Observing run &Exposure time &RV & N & V~$\sin$~i & Spectroscopic membership\\
P95 & & &(CASPEC)&[s]&[km s$^{-1}$] & & [km s$^{-1}$] &(CORAVEL) \\
\hline
 R10A & 12.99  & 1.04&July 1996&3600 &--&--&   -- &  No information\\
 R14 &  12.06& 	 0.60&April 1995&1200 & $-$15.60$\pm$0.24 & 4 &  6.0$\pm$1.6 & M\\
R15A  & 12.28 &   0.65&July 1996&2700& $+$18.60$\pm$0.56 & 1 & 15.0$\pm$2.0 &   N?,SB \\
 R16A & 11.69& 	 0.66&April 1994&720 &$-$13.77$\pm$0.34 & 2 &  5.8$\pm$2.4  & M\\
 R26A & 14.89& 	 1.44&July 1996&5400 & --&--&-- & 	No information\\     
 R27 &  12.12& 	 0.62&April 1994&1200 & $-$14.98$\pm$0.33 & 3 &  9.8$\pm$1.3 & M\\
R35	& 14.54 &   1.25&July 1996&4800&   --&--&--&   No information\\
 R39A&  12.36& 	 0.65&April 1995&1320 &$-$14.03$\pm$0.23 & 4 &  4.5$\pm$1.5  & M	\\
 R42 &  11.83& 	 0.57&July 1996&1800 & $-$11.23$\pm$0.43 & 2 & 14.3$\pm$2.3 & M,SB 	\\
 R48A & 14.73& 	 1.22&July 1996&5400 & --&--&-- & 	No information\\
 R49A & 12.37& 	 0.63&July 1996&2700 &$-$13.90$\pm$0.50 & 1 &  5.9$\pm$2.8 & M	\\
 R51A & 12.25& 	 0.63&April 1995&1200 & $-$11.81$\pm$1.06 & 1 & 14.3$\pm$3.3 & M\\
 R53 &  13.40& 	 0.84&April 1995&1800 & $-$12.83$\pm$0.56 & 1 &  3.7$\pm$3.2 & M\\
 R55A&  12.33& 	 0.77&April 1995&1200 &  $-$19.58$\pm$0.47 & 1 &  9.0$\pm$1.8 & M,SB? \\
 R55B&  11.70& 	 0.54&July 1996&2400 &  $-$13.59$\pm$0.59 & 4 & 13.7$\pm$4.6 & M,SB\\
 R64  & 11.96& 	 0.67&April 1994&1200 & $-$14.44$\pm$0.56 & 2 &  5.8$\pm$1.8 & M	\\
 R66  & 12.78& 	 0.72&April 1995&1500 & $-$15.71$\pm$0.45 & 3 &  7.7$\pm$1.3 & M,SB2\\
 R73A & 10.95& 	 0.51&April 1994&600 &$-$ 12.56$\pm$1.11 & 4 & 18.7$\pm$2.2 & M,SB	\\
 R79A & 11.92& 	 0.70&July 1996&1800 & $-$15.71$\pm$0.33 & 2 &  6.9$\pm$1.5  & M	\\
 R82 &  12.87& 	 0.67&April 1995&1620 & $-$14.69$\pm$0.30 & 3 &  6.7$\pm$1.5 & M	\\
 R92 &  12.46& 	 0.63&April 1995&1020 &  $+$4.31$\pm$1.03 & 2 &  8.1$\pm$1.6 & M,SB?\\
 R97 &  12.17& 	 0.63&April 1994&900 & $-$13.97$\pm$0.50 & 1 &  9.7$\pm$1.8 & 	M \\
 R102&  13.32& 	 0.83&April 1995&2100 & $-$15.97$\pm$0.56 & 1 &  6.1$\pm$4.3 & M\\
 R103 & 12.37& 	 0.65&April 1995&1500 & $-$13.56$\pm$0.43 & 1 &  2.4$\pm$2.8 & M	\\
%R104	& 12.37 &   0.76&  --&   M,Li not detectable\\
 R105 & 12.26& 	 0.62&April 1994&1020 & $-$10.22$\pm$0.53 & 1 & 11.0$\pm$2.0 & M,SB	\\
 R109A& 12.56& 	 0.71&April 1994&1200 &  $-$6.31$\pm$1.06 & 2 &  6.8$\pm$3.0 & M?,SB    \\
 R116&  12.77& 	 0.82&April 1995&1560 &  $+$10.59$\pm$25.86 & 3 &  7.5$\pm$1.8  & M?,SB	\\
 R123 & 13.17& 	 0.79&April 1995&1800 & $-$15.37$\pm$0.54 & 1 &  9.8 $\pm$2.1 & M\\
 R126A& 11.45& 	 0.50&April 1994&720 &--&--&-- & No information	\\
 R133 & 12.17& 	 0.65&April 1995&1200 &  $-$12.72$\pm$16.07 & 3 &  7.5$\pm$1.3	& M?,SB\\
 R135A& 13.08& 	 0.96&July 1996&3300 &--&--&--& 	No information\\
 R136A& 11.43& 	 0.70&July 1996&1800 & $-$14.83$\pm$0.81 & 1 & 26.2$\pm$2.6 & M,SB2	 \\
 R137A& 13.43& 	 0.93&July 1996&3600 & --&--&	-- &No information\\
 R140B& 12.38& 	 0.76&July 1996&3000 &  $-$14.83$\pm$0.44 & 1 &  1.9$\pm$3.1 & M	\\
%R141A	& 10.97 &   ?   &   --&   No information, Li not detectable\\

\hline
\end{tabular}
\end{table*}
%check R92,R123,R126A
%
\begin{table*}[!ht] \footnotesize
\caption{Radial and rotational velocities 
and membership for part of the stars of the James \& Jeffries
(1997) sample.}\label{memberJJ97}
\begin{tabular}{ccccc}
\hline
\hline
no. & RV & N & V~$\sin$~i & Spectroscopic membership \\
JJ97 & [km s$^{-1}$] & & [km s$^{-1}$] &(CORAVEL) \\
\hline
1 &$-$15.79$\pm$0.33 & 3 & 11.6$\pm$1.1 & M \\
3  &$-$15.57$\pm$2.36 & 2 & 65.8$\pm$13.4& M \\
6  &$-$14.49$\pm$0.50 & 3 & 19.4$\pm$1.2 & M \\
7  &$-$19.15$\pm$0.45 & 2 & 12.0$\pm$1.5 & M,SB \\
8  &$-$14.74$\pm$0.54 & 1 &  4.3$\pm$2.9& M \\
19 & $-$12.71$\pm$0.47 & 1 &  5.2$\pm$3.1& M \\
22  & $-$15.37$\pm$0.53 & 3 & 17.3$\pm$1.5& M \\
24  &$-$13.24$\pm$0.53 & 1 &  7.8$\pm$2.9 & M \\
27  &$-$11.19$\pm$0.59 & 1 & 15.5$\pm$2.1 & M,SB2 \\
29  & $-$13.45$\pm$0.65 & 3 &  4.2$\pm$1.9& M,SB? \\
31 & $+$34.44$\pm$7.37 & 2 & 19.2$\pm$1.3 & M,SB \\
33 & $-$14.91$\pm$0.38 & 1 & $<$3.2   & M\\
34 & $-$6.31$\pm$1.96 & 2 &  6.8$\pm$3.0 & M.SB \\
36&$-$24.67$\pm$3.24 & 2 & 12.6$\pm$2.4 & M,SB \\
%40 &--&--&-- & M?\\
%41 &--&--&--  & M?\\
%42 &--&--&--  & M?\\
\hline
\end{tabular}
\end{table*}
%$\mathrm{EW({\lambda}~6708~\AA)_{corr}}$
\section{Analysis}
\subsection{Radial velocities and membership}\label{RVanalysis}
The radial velocities are on the system defined by Udry et
al.~(\cite{umq}) from 
high--precision radial--velocities obtained with the ELODIE spectrograph 
(Baranne et al.~\cite{bqm}). This calibration corrects for the systematic 
errors of the 
CORAVEL system.
V~$\sin$~i values are derived from the width of the cross-correlation 
function according to the calibration of Benz \& Mayor (\cite{bm84}).
Our mean results for individual stars in NGC~6475 are summarized in 
Cols.~6--8 of
Table~\ref{original} (our sample) and Cols.~2--4 of Table~\ref{memberJJ97}
(JJ97 sample) which give the 
mean radial velocities with the errors in [km~s$^{-1}$], the 
number of radial--velocity observations, and 
the projected rotational velocities 
V~$\sin$~i and their errors, also in [km~s$^{-1}$].
  
Results for further stars and 
individual observations will be discussed in a separate paper devoted to the 
study of NGC~6475 based on CORAVEL observations. If necessary, data used 
in the present paper are available from J.--C. Mermilliod. 

CORAVEL radial velocities have been used in conjunction with available data,
from P95 and JJ97, which are usually based on one observation, 
to detect spectroscopic binaries and confirm the membership of the other 
stars. The results are recorded in Col.~9 of
Table~\ref{original} and Col.~5 of Table~\ref{memberJJ97}
as membership determinations and remarks on duplicity
(M: member, M?: possible member, N?: doubtful member, 
N: non--member, SB: spectroscopic binary, SB?: possible
spectroscopic binary, No information: the
star has not been observed with CORAVEL). In Table~\ref{memberJJ97}
the identification numbers of JJ97 are used (Col.~1).

Several stars require a comment:

\begin{itemize}
\item R15A: it is a binary, but all three observations are positive. The 
membership is doubtful.
\item R42: the difference between the RV of JJ97 ($-$16.2 km~s$^{-1}$) and 
CORAVEL RV reaches 
5 km~s$^{-1}$ and, according to the errors, R42 is certainly a binary.
\item R55A: the only RV obtained is off the cluster mean velocity by 
4 km~s$^{-1}$ and 
should be a binary if it is a member, as judged from it position in the CMD.
\item R66: it has been declared SB2 by P95 and its radial velocity is clearly 
variable.
\item R92: this star is either a binary or a non--member. The two RVs 
differ by 2 km~s$^{-1}$ and it may be variable.
\item R105: the difference between the published RV ($-$16.0 km~s$^{-1}$, 
P95) and CORAVEL data ($-$10.2 km~s$^{-1}$) supports binarity.
\item JJ7: there is a difference of 4 km~s$^{-1}$ between published and CORAVEL data.
\item JJ27: JJ97 and present data agree on binarity.
\item JJ31: large amplitude binary (JJ97 and present data).
\item JJ34: binary based on all data (P95, JJ97 and present data).
\item JJ36: clearly binary from CORAVEL and JJ97 data.
\end{itemize}

%The membership of JJ40, JJ41, JJ42 requires confirmation because the available 
%velocities are based on a single observation and differ by more than 
%18 km s$^{-1}$ from the cluster mean. If member, they clearly should be  
%binaries.

The cluster mean velocity, $-$14.63$\pm$0.18 (s.e. 0.87) km~s$^{-1}$, has
been computed from 24 stars not recorded as spectroscopic binaries.

We end this section with two comments: first, the high rate of confirmed members shows that X--ray surveys are effective in detecting new cluster members,
not only for very young clusters like \object{IC~2602} and \object{IC~2391}
(e.g. Randich et al.~\cite{2602}),
but also
for somewhat older clusters; second, whereas
a detailed discussion of the evolution of rotation is beyond the
scope of this paper, we note that
the projected rotational velocities are rather low 
for the majority of the stars. More specifically,
considering our sample (Table~\ref{original}) and the stars
from the sample of JJ97 (Table~\ref{memberJJ97}),
there are 26 stars ($\rm{\sim\,63\, \%}$) 
with $\rm{V \sin i \leq{10}}$ km~s$^{-1}$,
13 stars ($\rm{\sim\,32\, \%}$) with $\rm{10<{V \sin i}\leq{20}}$ km~s$^{-1}$
and only 2 stars ($\rm{\sim\,5\, \%}$) with $\rm{V \sin i >{20}}$ km~s$^{-1}$:
star R136A
($\rm{V \sin i =26.2\pm2.6}$ km~s$^{-1}$) and star JJ3 
($\rm{V \sin i =65.8\pm13.4}$km~s$^{-1}$).

\begin{table}[!ht] \footnotesize
\caption{Li equivalent widths and Li abundances for the stars in our sample.
For stars warmer than 4500~K Li abundances
are corrected for NLTE effects; for the coolest stars
the LTE Li abundances are reported. Star R26A, marked with an asterisk,
is discussed in Sect.~\ref{Liabundance}.}\label{tab6475}
\begin{tabular}{ccccc}
\hline
\hline
no. & \teff & $\mathrm{EW(Li+Fe)}$& $\mathrm{EW(Li)}$ & \nli\\
P95 &      [$\mathrm{K}$] & [$\mathrm{m\AA}$] & [$\mathrm{m\AA}$] &\\
\hline
 R10A & 4507	&	 153$\pm$5&135$\pm$5 	&	 1.63$\pm$0.15 \\
 R14 &5888	&	 107$\pm$6&98$\pm$6 	&	 2.71$\pm$0.10 \\
 R16A &	5656	&	 155$\pm$10&145$\pm$10 	&	 2.77$\pm$0.17 \\
 R26A & 3860	&	 $<30$* &$<30$ 	&	 $<-0.20$\\
 R27	&5810	&	 122$\pm$3&113$\pm$3 	&	 2.73$\pm$0.10 \\
R35    &4095   &        73$\pm$18&51$\pm$18    &        0.28$\pm$0.21\\
 R39A	&5696	&	 108$\pm$5&98$\pm$5 	&	 2.56$\pm$0.10 \\
 R42  	&6008	&	 115$\pm$4&107$\pm$4 	&	 2.85$\pm$0.10 \\
 R48A 	&4144	&	 41$\pm$10&20$\pm$10 	&	 $-$0.12$\pm$0.28\\
 R49A 	&5772	&	 92$\pm$6&82$\pm$6 	&	 2.53$\pm$0.10 \\
 R51A	&5772	&	 46$\pm$5&36$\pm$5 	&	 2.13$\pm$0.11 \\
 R53 	&5048	&	 88$\pm$12&74$\pm$12 	&	 1.84$\pm$0.15 \\
 R55A	&5272	&	 86$\pm$10&74$\pm$10 	&	 2.05$\pm$0.14 \\
 R55B	&6131	&	 91$\pm$3&83$\pm$3 	&	 2.80$\pm$0.09 \\
 R64 	&5622	&	 77$\pm$5&67$\pm$5 	&	 2.30$\pm$0.11 \\
 R66   	&5442	&	 139$\pm$5&128$\pm$5 	&	 2.51$\pm$0.11 \\
 R73A 	&6257	&	 63$\pm$15&56$\pm$15 	&	 2.69$\pm$0.17 \\
 R79A 	&5513	&	 142$\pm$8&131$\pm$8 	&	 2.59$\pm$0.12 \\
 R82 	&5622	&	 128$\pm$5&118$\pm$5 	&	 2.61$\pm$0.10 \\
 R92	&5772	&	 110$\pm$5&100$\pm$5 	&	 2.64$\pm$0.10 \\
 R97 	&5772	&	 120$\pm$8&110$\pm$8 	&	 2.69$\pm$0.11 \\
 R102   &5079	&	 81$\pm$5&67$\pm$5 	&	 1.82$\pm$0.13 \\
 R103	&5696	&	 111$\pm$5&101$\pm$5 	&	 2.58$\pm$0.10 \\
 R105	&5810	&	 115$\pm$3&106$\pm$3 	&	 2.70$\pm$0.10 \\
 R109A  &5477	&	 121$\pm$7&110$\pm$7 	&	 2.43$\pm$0.12 \\
 R116 	&5110	&	 83$\pm$4&70$\pm$4 	&	 1.87$\pm$0.12 \\
 R123 	&5206	&	 107$\pm$6&94$\pm$6 	&	 2.12$\pm$0.12 \\
 R126A	&6300	&	 100$\pm$6&93$\pm$6 	&	 2.98$\pm$0.09 \\
 R133 	&5696	&	 145$\pm$5&135$\pm$5 	&	 2.76$\pm$0.10 \\
 R135A	&4706	&	 170$\pm$3&154$\pm$3 	&	 1.76$\pm$0.14 \\
 R136A	&5513	&	 108$\pm$6&97$\pm$6 	&	 2.41$\pm$0.11 \\
 R137A	&4787	&	 156$\pm$3&140$\pm$3 	&	 1.79$\pm$0.13 \\
 R140B	&5305	&	 105$\pm$3&93$\pm$3 	&	 2.20$\pm$0.11  \\

\hline
\end{tabular}
\end{table}
\begin{table*}[!ht] \footnotesize
\caption{Stellar parameters and Li abundances for the sample of
JJ97. Asterisks denote stars in common with our sample.
For stars warmer than 4500~K Li abundances
are corrected for NLTE effects; for the coolest stars
the LTE Li abundances are reported.}\label{tabJJ97} 
\begin{tabular}{ccccccccccc}
\hline
\hline
no. & no. & V &$\mathrm{(B-V)_{0}}$ & \teff & $\mathrm{EW(Li)}$ &\nli\\
P95 & JJ &    & & [$\mathrm{K}$] & [$\mathrm{m\AA}$] & \\
\hline
61&  1 &	 11.58 & 0.62 & 5810	 & 124$\pm$4  & 2.79$\pm$0.10\\
42&  2* &	 11.38 &0.57 & 6008	 & 105$\pm$9  &	2.84$\pm$0.11\\
81&  3 &	 11.38 & 0.48 & 6386	 &  58$\pm$13  &2.80$\pm$0.15\\
69&  4 &	 12.13 & 0.60 & 5888	 &  30$\pm$10  &2.14$\pm$0.33\\
127B&  6 &	 11.67 &0.57 & 6008	 &  90$\pm$14 &	  2.75$\pm$0.13\\
127A&  7 &	 11.88 &0.61 & 6330   & 101$\pm$9  &3.05$\pm$0.10\\
94&  8 &	 13.30 & 0.81 & 5142   & 126$\pm$10  &2.23$\pm$0.13\\
82&  9* &	 12.87 &0.67 & 5622	 & 127$\pm$11 & 2.66$\pm$0.12 \\
82B& 10 &	 12.53 &0.68 & 5585	 & 109$\pm$9  &	  2.54$\pm$0.12 \\
53& 11* &	 13.40 & 0.84 & 5048	 &  68$\pm$12  &1.79$\pm$0.16\\
27& 12* &	 12.12 &0.62 & 5810	 & 101$\pm$8  &	  2.67$\pm$0.11\\
16A& 13* &	 11.69 &0.66 & 5659	 & 102$\pm$7  &	  2.56$\pm$0.11\\
7B& 14 &	 13.78 & 1.02 & 4555	 & 123$\pm$17  &1.62$\pm$0.17\\
7A& 15 &	 14.15& 1.00 & 4604	 & 203$\pm$25  &2.04$\pm$0.21\\
%--& 16 &	 12.36& 0.65 & && 5696& 116$\pm$10  &2.66$\pm$0.12 & --\\
39A& 16* &	 12.36 &0.65 & 5696	 & 134$\pm$13 &	  2.75$\pm$0.13\\
--& 17 &	 12.30 & 0.69 & 5549	 &  91$\pm$11  &2.40$\pm$0.13\\
--& 18 &	 11.99 & 0.71 & 5477	 & 124$\pm$10  &2.52$\pm$0.12\\
76& 19 &	 13.55 & 0.89 & 4899	 & 104$\pm$11  &1.88$\pm$0.14\\
--& 20 &	 14.05& 0.98 & 4654	 &  59$\pm$12  &1.32$\pm$0.17\\
24& 22 &	 11.11 & 0.44 & 6564	 &  60$\pm$8  & 2.90$\pm$0.15\\
103& 23* &	 12.37 &0.65 &5696	 &  99$\pm$12 &	  2.57$\pm$0.13\\
33& 24 &	 13.42 & 0.83 & 5079	 & 127$\pm$18  &2.18$\pm$0.16\\
102& 25* &	 13.32 &0.83 & 5079   & 108$\pm$34 & 2.08$\pm$0.25\\
14& 26* &	 12.06 &0.60 &5888	 &  97$\pm$6  &	  2.70$\pm$0.10\\
104& 27 &	 12.37 &0.76 & 5305	 & 126$\pm$8  &	  2.38$\pm$0.12\\
72& 28 &	 10.79 & 0.49 & 6343	 &  59$\pm$8  & 2.77$\pm$0.11\\
95& 29 &	 12.19 & 0.62 & 5810	 &  97$\pm$8  & 2.64$\pm$0.11\\
39B& 31 &	 12.19 &0.63 &5772   & 118$\pm$11 & 2.73$\pm$0.12 \\
119A& 33 &	 12.92 &0.75 & 5339	 & 103$\pm$9  &	  2.29$\pm$0.12\\
109& 34 &	 12.63 &0.74 & 5373	 &  90$\pm$8  &	  2.25$\pm$0.13\\
66& 35* &	 12.78 &0.72 & 5442	 &  87$\pm$7  &	  2.29$\pm$0.12\\
132& 36 &	 11.81 &0.62 & 5810	 &  49$\pm$14 & 2.29$\pm$0.17\\
1& 40 &	         13.30 &1.07 & 4439	 &  17$\pm$9  &	  1.84$\pm$0.31\\
--& 41 &	 11.57 & 0.52 & 6214	 &  89$\pm$8  & 2.89$\pm$0.10 \\
134& 42 &	 12.41 &0.78 & 5239	 &  13$\pm$4  &	  1.21$\pm$0.18\\
\hline
\end{tabular}
\end{table*}
\subsection{Abundances analysis}\label{abundance}
Effective temperatures (listed in Col.~2
of Table \ref{tab6475}) were computed from dereddened \bmv~colors,
and using the calibration of S93: namely,
$\rm{T_{eff}=1808{(B-V)_{0}}^{2}-6103(B-V)_{0}+8899}$ K;
we assume conservative random uncertainties $\rm{\Delta{T_{eff}}=\pm100}$ K.

We assumed for all the sample stars
the same surface gravity 
$\log g=4.5$, while microturbulence
was derived as $\xi=3.2 \times 10^{-4} (\rm T_{\rm eff}-6390)-1.3(\log g-4.16)+1.7$ (Nissen \cite{nissen},
Boesgaard \& Friel  \cite{bf90}); these two parameters have
 negligible effects
on Li abundances, while they affect metallicity. The assumed random errors
are 0.3 dex in $\log g$ and 0.3 km~s$^{-1}$ in $\xi$.
\subsubsection{Lithium}\label{Liabundance}
We carried out the Li analysis for 33 stars in our sample:
star R15A was excluded since it is a doubtful member; 
the stars with no CORAVEL information on membership (see Table~\ref{original})
were considered as possible members (M?), given their spectral
characteristics and Li abundances.

We measured equivalent widths
(EWs) of the Li~{\sc i}~$\mathrm{{\lambda}6707.8\, \AA}$ doublet; at our
resolutions, this spectral feature is blended, 
or partially blended, with the
Fe~{\sc i}~$\mathrm{{\lambda}6707.44\, \AA}$ line, therefore the
contribution of the latter feature needs to be considered.
The EW of the iron line was estimated
following the prescription of S93, namely
EW(Fe)$\mathrm{=[20(B-V)_{0}-3]\,m\AA}$.
Note that the EW of star R26A
(\teff$=$3860~K), marked with one asterisk in Col.~3 of Table~\ref{tab6475},
has not been corrected
for the Fe~{\sc i} contribution, since the
S93 formula is no more valid for \teff~below 4000~K (see discussion
in Randich et al.~\cite{R00}). 
In this case, the quoted \nli~should be regarded as
an upper limit.

Li abundances (\nli) were computed from the measured EWs by interpolating
the curves of growth (COG) of S93; Li abundances were then corrected for
non local thermodynamic equilibrium (NLTE) effects by using the code
of Carlsson et al. (\cite{carl94}): NLTE corrections are provided only
down to~\teff$=$4500~K. Whereas for cooler stars we adopted
LTE Li abundances, NLTE corrections are small
below 4500~K
(e.g. Pavlenko et al.~\cite{pav}).
The measured EWs of the Li$+$Fe feature and the 
EWs corrected for the Fe~{\sc i} blend
are listed in Cols.~3 and~4 of Table~\ref{tab6475}, while
the derived Li abundances are listed in Col.~5.
Uncertainties in \nli~were computed by quadratically adding
the errors due to uncertainties in~\teff~and in EWs. 

In the following, we will compare our results 
with those of JJ97 and J00
for NGC~6475 and with those of other clusters.
In order to put all the data on a homogeneous 
scale, we recomputed Li abundances
using the procedure described above and starting from published EWs
for these clusters
(NGC~6475 --JJ97 and J00;
M~34 --Jones et al.~\cite{jones97};
Pleiades --S93 and Jones et al.~\cite{jon96};
and Hyades --Thorburn et al.~\cite{thor93}).
Where necessary we also recomputed
the effective temperatures (i.e. when the \teff~vs.~\bmv~calibrations
used by other authors were different from the \teff~used here).

Stars, photometry, EWs and \nli~for the samples of JJ97 and J00
are listed in Tables~\ref{tabJJ97} and \ref{tabJ00}.
The \teff~listed in in Tables~\ref{tabJJ97} and \ref{tabJ00}
are those recomputed by us, after
retrieving \bmv~values from P95.
The listed EWs were corrected for
the Fe~{\sc i} blend by JJ97 and J00, using a spectral subtraction technique.
Ten of the stars of JJ97 (marked with asterisks
in Table~\ref{tabJJ97}) are in common with our sample: for this reason,
in the first two
columns of Table~\ref{tabJJ97}, 
we report both the star numbers of P95 and of JJ97, while
in Table~\ref{tabJ00} only the JJ97 numbering system is adopted.

All the stars in Table~\ref{tabJJ97}, 
with the exception of JJ40, JJ41 and JJ42,
were considered as \emph{bona fide} cluster members (see JJ97 and Table~\ref{memberJJ97}).
The membership of JJ40, JJ41, JJ42 requires confirmation because the available 
velocities are based on a single observation and differ by more than 
18 km~s$^{-1}$ from the cluster mean. If members, they clearly should be  
binaries.

Stars in Table~\ref{tabJ00} were selected by us following the same criteria
adopted by J00, i.e. stars with radial velocities far from
the cluster mean velocity were rejected; 
since star JJ105, rejected
by J00, has a RV differing only by 5 km~s$^{-1}$ from
the cluster mean velocity, we consider this stars as a possible member.
\subsubsection{Metallicity}\label{metal}
\begin{table*}[!ht] \footnotesize
\caption{Stellar parameters and Li abundances for the sample of
James et al.~2000. Li abundances are corrected for NLTE effects.}\label{tabJ00}
\begin{tabular}{cccccc}
\hline
\hline
no. & V &$\mathrm{(B-V)_{0}}$ & \teff & $\mathrm{EW(Li)}$ & \nli\\
JJ &    & & [$\mathrm{K}$] & [$\mathrm{m\AA}$] & \\
\hline
 103	& 12.55 & 	 0.66 &  5658 	&123$\pm$15  & 	 2.66$\pm$0.14 \\
 104	& 12.18 & 	 0.72 &  5442 	&192$\pm$13  & 	 2.80$\pm$0.14 \\
 105	& 12.92 & 	 0.77 &  5272 	& 34$\pm$15  & 	 1.67$\pm$0.25 \\
 106	& 12.20 & 	 0.64 &  5734 	&148$\pm$11  & 	 2.84$\pm$0.12 \\
 109	& 12.07 & 	 0.67 &  5622 	&156$\pm$35  & 	 2.79$\pm$0.24 \\
 110	& 12.96 & 	 0.80 &  5174 	& 90$\pm$13  & 	 2.06$\pm$0.15 \\
 111	& 11.59 & 	 0.52 &  6214 	&132$\pm$13  & 	 3.12$\pm$0.12 \\
 113	& 11.80 & 	 0.59 &  5928 	&133$\pm$17  & 	 2.92$\pm$0.14 \\
 115	& 11.70 & 	 0.60 &  5888 	&137$\pm$13  & 	 2.91$\pm$0.12 \\
 116	& 12.29 & 	 0.64 &  5734 	&158$\pm$12  & 	 2.88$\pm$0.12 \\
\hline
\end{tabular}
\end{table*}
\begin{table*}[!ht] \footnotesize
\caption{Iron abundances.}\label{iron}
\begin{tabular}{lccccccccc} \hline
\hline
Star &  \multicolumn{9}{c}{$\log$~n(Fe)} \\
     & 6703.57~\AA & 6725.36~\AA & 6726.67~\AA & 6750.16~\AA & 6810.27~\AA & 6820.37~\AA & 6828.60~\AA & 6858.15~\AA & average($\pm\sigma_1\pm\sigma_2$)\\ 
\hline
R14 &  -- & 7.87 & 7.61 & -- & 7.56 & 7.67 & 7.65 & 7.48 & 7.64$\pm$0.13$\pm$0.11\\
R16A&  7.64 & -- & 7.65 & 7.64 & 7.67 & 7.72 & 7.72 & 7.50 & 7.65$\pm$0.07$\pm$0.11\\
R39A&  7.65 & 7.76 & 7.66 & 7.57 & 7.52 & 7.64 & 7.61 & -- & 7.63$\pm$0.08$\pm$0.11\\
R49A&  7.62 & -- & -- & 7.55 & 7.89 & 7.56 & -- & 7.42 & 7.61$\pm$0.17$\pm$0.11\\
R55A&  7.68 & 7.80 & 7.66 & 7.75 & 7.68 & 7.73 & -- & 7.62 & 7.70$\pm$0.06$\pm$0.11\\
R66 &  7.61 & -- & 7.56 & 7.81 & 7.60 & 7.60 & 7.58 & 7.49 & 7.61$\pm$0.10$\pm$0.11\\
R92 &  7.74 & -- & 7.68 & 7.78 & 7.67 & 7.69 & 7.66 & 7.53 & 7.68$\pm$0.08$\pm$0.11\\
R102&  7.63 & -- & 7.79 & 7.77 & 7.61 & 7.58 & 7.61 & 7.47 & 7.64$\pm$0.11$\pm$0.14\\
R103&  7.69 & 7.64 & 7.83 & 7.59 & -- & 7.64 & 7.73 & 7.63 & 7.68$\pm$0.08$\pm$0.11\\
R105&  -- & 7.67 & 7.74 & -- & 7.67 & -- & 7.61 & 7.61 & 7.66$\pm$0.06$\pm$0.11\\
R123&  7.55 & -- & -- & 7.76 & -- & 7.70 & 7.68 & 7.67 & 7.67$\pm$0.08$\pm$0.11\\
vB182&  7.62 & 7.65 & 7.64 & 7.63 & -- & -- & -- & -- & 7.64$\pm$0.02$\pm$0.10\\
vB187&  7.66 & 7.67 & 7.68 & 7.64 & -- & -- & -- & -- & 7.66$\pm$0.02$\pm$0.10\\
\hline
\end{tabular}
\end{table*}

We used the best quality spectra to derive the cluster metallicity.
The iron abundance analysis was carried out using MOOG (Sneden
\cite{sneden}--version December 2000) and Kurucz (\cite{kuru})
model atmospheres. 
For each star we measured up to eight
Fe~{\sc i} lines, whose wavelengths are listed in Table~\ref{iron}
(the EWs are available from S.~Randich, upon request); 
for these lines we adjusted $\log gf$ values by carrying 
out an inverse abundance analysis
of the solar spectrum. We used the spectrum of the Sun observed
with the FEROS instrument at La Silla, during another observing run.
The resolving power of the FEROS spectrum (R$\sim$48,000) is somewhat
higher than that of our sample spectra; all the lines that we used for the 
iron analysis, however, do not have close features that could be blended
at our resolution and that could lead to overestimate the cluster metallicity.
For the Sun, we assumed $\log$~n(Fe)$=7.52$ and the usual solar parameters:
T$_{\rm eff\odot}=5770$~K, $\log g_{\odot}=4.44$, and $\xi_{\odot}=1.1$~
km~s$^{-1}$.
Van der Waals broadening was treated using the Uns\"old
approximation (\cite{uns55}).

The derived iron abundances for each line and the mean abundance for
each star are listed in Table~\ref{iron}; the random errors in the mean, $\sigma_1$
and $\sigma_2$ are also listed in the table.
The random errors were estimated similarly to Randich et
al.~(\cite{R00}); namely, we assumed that, for each star,
the standard deviation of the mean iron abundance would be a good
estimate of the error --$\sigma_1$-- due to errors in measured EWs and to 
random
uncertainties in atomic parameters, $gf$--values in particular. We then
estimated random errors due to uncertainties in stellar parameters
--$\sigma_2$-- by 
varying each parameter at a time and leaving the other two parameters
unchanged; we then quadratically added
the related errors. As mentioned above,
conservative errors of 100~K in T$_{\rm eff}$, 0.3~dex
in $\log g$, and 0.3 km~s$^{-1}$ in $\xi$ were assumed.
Note that we did not find any abundance trend vs. EW or EP
(excitation potential), meaning that
our assumed parameters should not be largely in error.

In order to estimate systematic errors and to put our [Fe/H] determination
for NGC~6475 on a consistent scale with other well known clusters,
we determined [Fe/H] for two Hyades members.
The two stars, vB182 and vB187, which have T$_{\rm eff}$ within the range
covered by our sample stars, were observed by us during one observing
run with UVES on VLT UT2 (the detailed description of the data, analysis,
and results will be reported elsewhere). The spectra have resolving power
R$\sim$40,000 and S/N ratios around 200. 
We derived stellar parameters for the two Hyades stars consistently with
our sample stars and obtained (or assumed in the case of surface gravity):
\teff$=$5079~K, $\log g=4.5$, $\xi=0.84$~km~s$^{-1}$ for vB182
and \teff$=$5339~K, $\log g=4.5$, $\xi=0.92$~km~s$^{-1}$ for vB187.
We measured the EWs of the four
Fe~{\sc i} lines of the ones used by us for metallicity
included in the UVES spectral range
and derived iron abundances as for our sample stars. The abundances for the
two Hyades stars are listed in Table~\ref{iron}.

We computed the weighted mean iron abundance for NGC~6475 using all the stars
listed in Table~\ref{iron} and obtained $\log$~n(Fe)
$=7.66\pm 0.06$, or [Fe/H]$\rm{=+0.14 \pm 0.06}$. Note that, when computing
the mean, we conservatively assumed for each star a total error
$\sigma=\sigma_1+\sigma_2$. The mean for the Hyades is $\log$~n(Fe)=
7.65 $\pm 0.08$ or [Fe/H]$=0.13 \pm 0.08$. In other words {\it a)}
our metallicity for the Hyades is virtually the same as the usually
quoted value for this cluster ([Fe/H]$=+0.13$, Boesgaard \& Budge
\cite{bb89}), implying that our analysis
should not be affected by large systematic errors; {\it b)} the metallicity
of NGC~6475 is over--solar and very similar to that of the Hyades.

As mentioned in the introduction a metallicity larger than solar
for NGC~6475 was already found by JJ97, who derived [Fe/H]$=+0.11 \pm 0.034$,
in good agreement with our estimate. In addition, one star
in the sample of JJ97 (JJ26/R14) is 
in common with the sample we have used for the
metallicity determination; 
the metallicity we derive for this star is consistent with
the value quoted by JJ97 ([Fe/H]=$+0.12$ and $+0.10$, respectively).
However, whereas JJ97 assumed the same \teff~values
and similar $\log g$ as ours, they assumed a microturbulence 
$\xi=2$~km~s$^{-1}$ for all the stars. Had we assumed this value, we would
have found a much lower metallicity for the cluster ([Fe/H] $\sim$ solar).
We note that our choice for the microturbulence parameter 
is more in agreement
with other metallicity studies: besides 
Boesgaard \& Friel (\cite{bf90}), several other authors used low, and
temperature dependent, values of $\xi$ for cluster and field dwarfs (e.g., 
Edvardsson et al.~\cite{ed93}; King et al.~\cite{king00}).
In addition, our microturbulence scale is consistent with the
most commonly used value for the solar microturbulence which we also
assumed for our inverse analysis of the solar spectrum (see above).

\begin{figure}
\psfig{figure=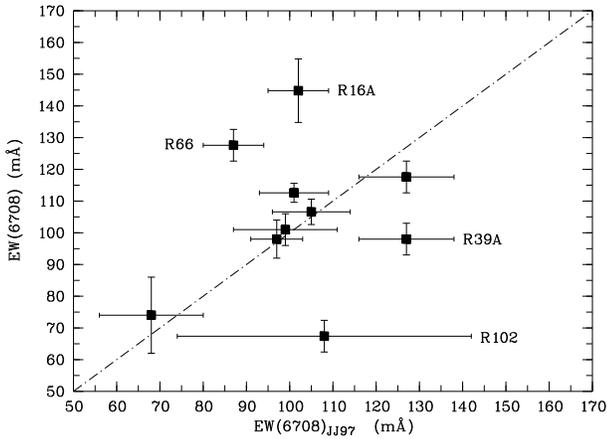, width=6.0cm, angle=-90}
\caption{Our NGC~6475 equivalent widths
are plotted vs.~the equivalent widths of
James \& Jeffries 1997
for the 10 stars in common.}\label{EW}
\end{figure}

\begin{figure}
\psfig{figure=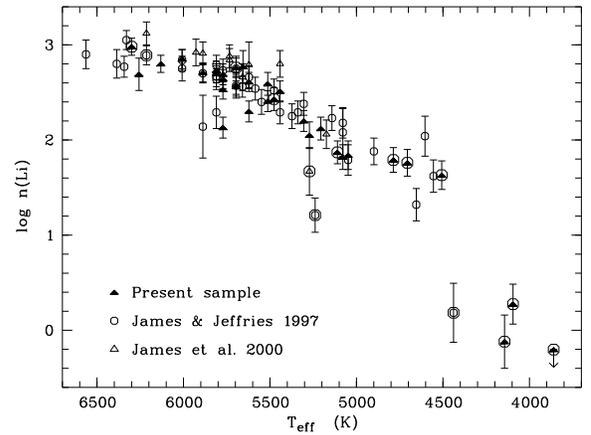, width=6.0cm, angle=-90}
\caption{NGC~6475 Li abundance vs.~\teff~:
the present sample (filled triangles) is compared
to those of James \& Jeffries 1997 (circles)
and James et al.~2000 (open triangles);
circled symbols
denote stars whose
membership is still to be confirmed, while down--pointing arrows
represent upper limits in~\nli.
Error bars are also shown.}\label{Li6475}
\end{figure}

\section{Results}\label{results}
\subsection{Lithium in NGC~6475}\label{Liresults}
In Fig.~\ref{EW} we plot our deblended Li EWs
vs.~the EWs measured by JJ97
for the ten stars in common.
The figure shows a good agreement between the EWs of JJ97
and our measurements; however four stars are present for which
the differences between the two measurements are larger than
the errors:
the discrepant stars are R16A and R66, for which our EWs are much larger 
than the JJ97 EWs, R39A and R102, which on the contrary lie
well below the mean trend.
The spectra of these stars (see Fig.~\ref{spectra})
have a rather high S/N ratio and the continuum can be
clearly determined, thus an error in our EW measurements is not likely;
moreover the [Fe/H] values obtained for these stars are
consistent with the average estimated metallicity, which means
that we have not systematically over/underestimated the EWs.
Note however that star R66 is a SB2 (see Table~\ref{original}): the discrepancy
between the two EW measurements might be due
to the fact that R66 was observed 
by us and JJ97 in two
different phases. As far as the other three discrepant stars
are concerned,
whereas there is no definitive answer
about the differences between the two sets of EWs,
we suggest that they
may be due to the different methods of subtraction
of the Fe~{\sc i}~EW 
(as mentioned, JJ97 used a spectral subtraction technique),
or to a lower S/N of the JJ97 spectra.

The \nli~vs.~\teff~ distributions for the three samples
are plotted in Fig.~\ref{Li6475}: 
the three distributions appear very
similar; in particular
no systematic difference (due, e.g., to different instruments,
spectral resolutions or reduction methods) is present;
the three sets therefore can be safely merged into
a single larger sample. 
We note that the differences between
our EWs and those of JJ97 for the ten stars in common
are strongly reduced when considering Li abundances;
in the following we will use our own Li measurements for the stars in common.
\begin{figure}
\psfig{figure=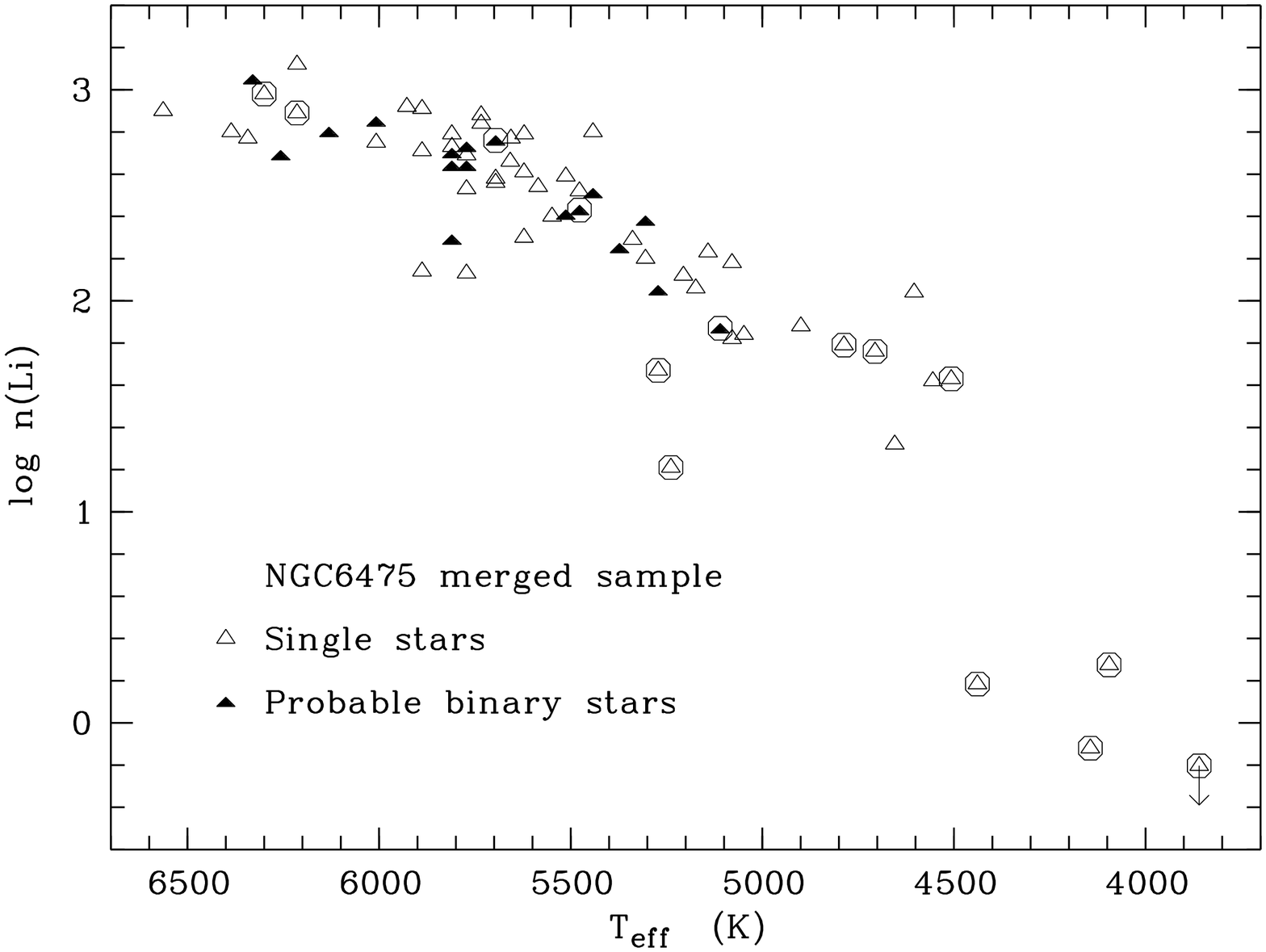, width=6.0cm, angle=-90}
\psfig{figure=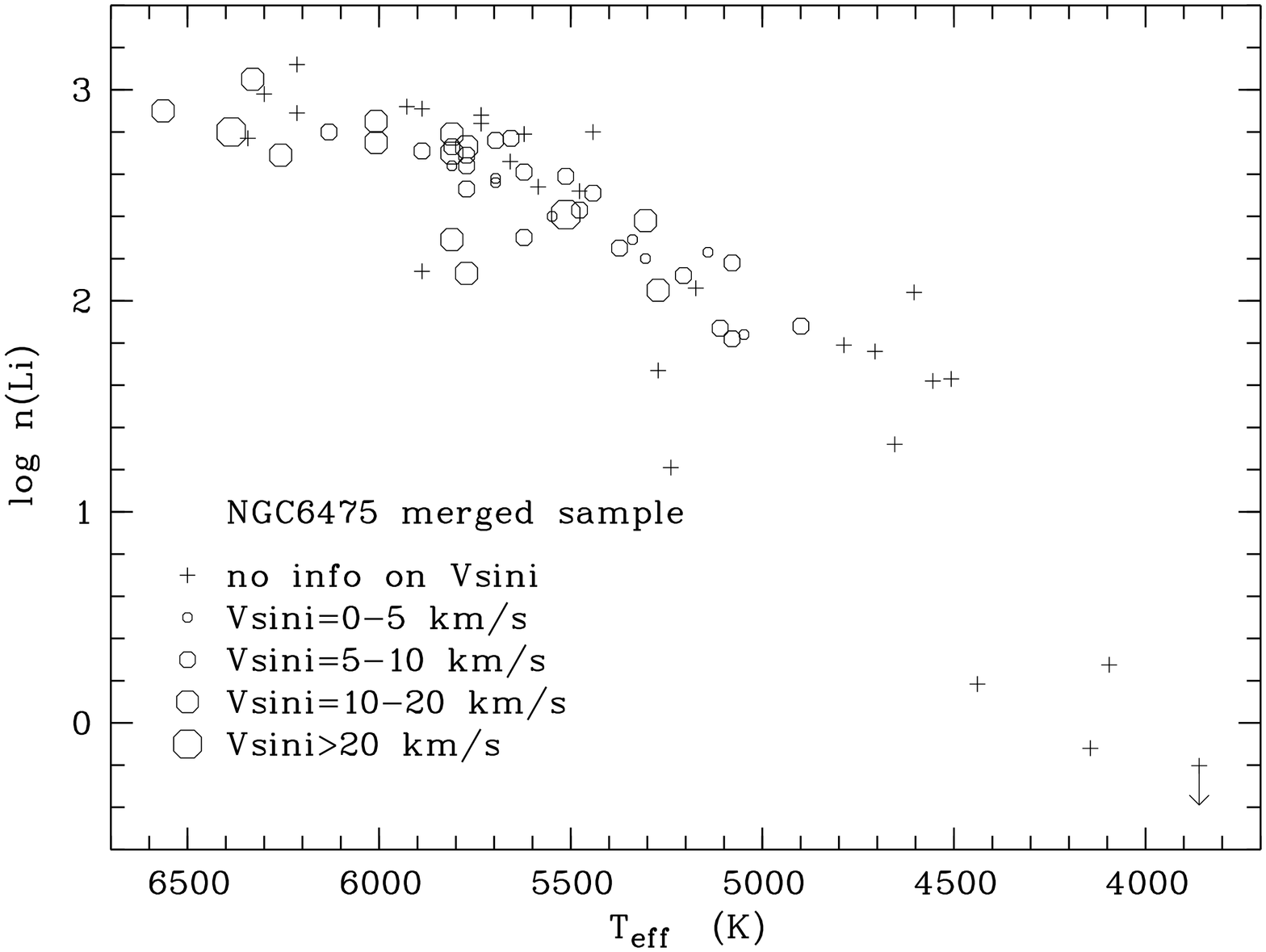, width=6.0cm, angle=-90}
\caption{Li abundance vs.~\teff~for the NGC~6475
merged sample (our sample+JJ97+J00).
Top panel: open triangles denote single stars, while filled triangles
indicate spectroscopic binary stars.
Circled symbols denote stars whose
membership is still to be confirmed.
Bottom panel: stars with available projected rotational velocity are
plotted as open circles: the size of the circles is proportional to
V~$\sin$~i; stars with no CORAVEL information on rotational velocity
are plotted as crosses.}\label{binarie}
\end{figure}

The Li vs.~\teff~distribution of NGC~6475 is almost
flat for the late F stars, showing little depletion
with respect to the initial value (\nli$_{0}=$3.1--3.3,
as indicated
from T Tauri stars and meteorites);
stars with \teff~below $\sim$ 6000~K have instead
undergone Li depletion with the Li distribution
showing a rapid decline as the stellar temperature
(mass) decreases.
No evident scatter
in Li abundances is present in this cluster for stars 
warmer than $\sim$ 4800~K, although a few stars
below the mean trend are present:
two of these stars have no confirmed membership
(JJ42, \teff$\rm{=}$5239 K and JJ105, \teff$\rm{=}$5272 K), one is
a probable spectroscopic binary (JJ36, \teff$\rm{=}$5810~K)
and two stars are confirmed as members
(JJ4, \teff$\rm{=}$5888 K and R51, \teff$\rm{=}$5772 K).
The Li spread among cooler stars will be discussed
later.

\begin{figure}
\psfig{figure=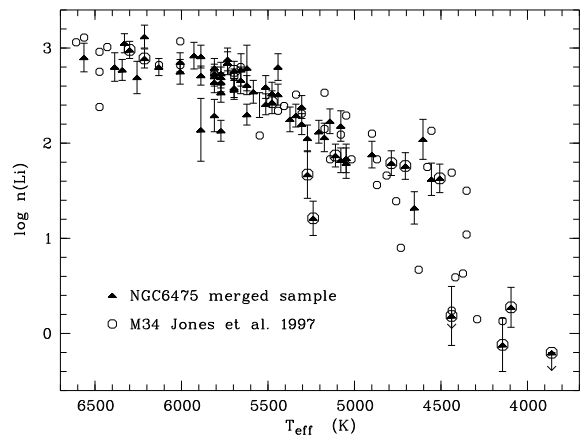, width=6.0cm, angle=-90}
\caption{Comparison of the Li distributions of the NGC~6475 merged sample
(filled triangles) and M~34 (Jones et al.~1997,
open circles); circled symbols
denote stars whose
membership is still to be confirmed.}\label{M34}
\end{figure}
In the top panel of Fig.~\ref{binarie} the NGC~6475 merged sample
(our stars+JJ97+J00) is shown:
single and probable spectroscopic binary stars are plotted
with different symbols.
This plot shows that, apart from JJ36 (see above) 
binarity character does not
alter significantly the determination of 
Li abundances because both kinds of stars 
are well mixed.

In the bottom panel of Fig.~\ref{binarie} NGC~6475 stars with
different projected rotational velocities
are represented with symbols of different
size: there is no evident correlation between Li abundances and V~$\sin$~i
values. 
We stress however that, with exception of two objects with
V~$\sin$~i$>$20 km~s$^{-1}$ (see Sect.~\ref{RVanalysis}),
all the other stars have low rotational velocities.
\subsection{Comparison with other clusters}
Fig~\ref{M34}  shows a comparison between the NGC~6475 merged sample
and M~34;
the two Li patterns are almost indistinguishable
in spite of the  slight
difference in metallicity, at least down to \teff$\sim$4800--5000~K.
The M~34 sample contains a significantly larger number of
mid and late K--type stars than ours.
An evident Li abundance star--to--star scatter is
present for \teff$\lesssim$4700--4800~K; Jones et al.~(\cite{jones97})
showed that this spread in M~34 cannot be attributed to
measurement uncertainties. We have some hints
of a similar scatter in NGC~6475, but our sample includes only very few
stars cooler than $\sim$ 4800~K and for most
of them the membership is uncertain.

\begin{figure}
\psfig{figure=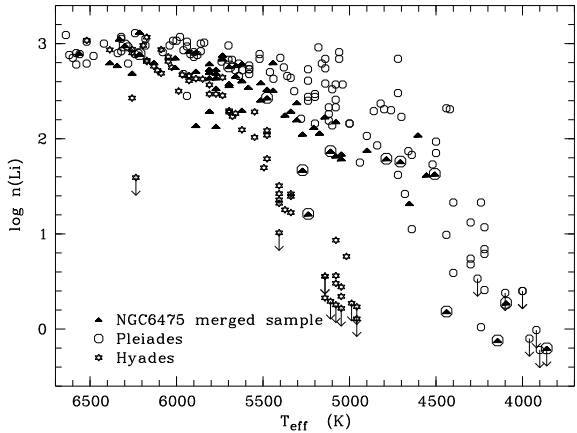, width=6.0cm, angle=-90}
\caption{The NGC~6475 Li distribution (filled triangles)
is compared to those of the Pleiades (S93$\rm{+}$Jones et al.~1996, circles) 
and the Hyades (Thorburn et al.~1993, stars). 
Circled symbols denote stars whose membership is
still to be confirmed.}\label{hyades}
\end{figure}

In Fig.~\ref{hyades} we show a comparison between the Li patterns
of
NGC~6475, the Pleiades and
the Hyades.
We can divide the plot in three temperature ranges:
{\it a)} \teff$\gesssim$6000~K: the three clusters
have similar Li distributions, with a mean value slightly
below the meteoritic abundance, i.e. these stars seem to suffer
a very little amount of MS Li depletion.
We do not consider here the Li dip observed in the Hyades stars
with \teff$\sim$6500~K, since our sample contains mostly stars
cooler than the dip.
The only Li poor star belonging to the Hyades 
has \teff$\sim$6200~K, but its membership is uncertain
and it is discussed in Thorburn et al.~\cite{thor93};
{\it b)} ${5500}\lesssim$\teff$\lesssim{6000}$ K: in this \teff~range, it is evident
that the NGC~6475 Li pattern lies between those of the Pleiades and the Hyades,
suggesting that Li depletion is a continuous process occurring for 
G--type stars
both between $\sim$ 100 and 220 Myr and between 220 and 600 Myr.
We stress that below $\sim$ 6000~K a few stars in NGC~6475 
appear as depleted as
(or more depleted than) the older Hyades:
these stars seem to be {\em bona fide} cluster members
(see Sect.~\ref{Liresults}) and should
be further monitored;
{\it c)} \teff$\lesssim$5500~K: as well known, Pleiades stars are characterized
by a large
amount of scatter for late G to late K stars.
The NGC~6475 distribution lies on the lower envelope of the
Pleiades distribution, and several Pleiades stars exist that show the same 
amount of depletion as NGC~6475;
the dispersion
is not present in NGC~6475 at least for stars hotter than $\sim$ 4800~K:
more precisely, above this temperature the only two stars
that could indicate the presence of a scatter,
as mentioned, are not confirmed as members;
this shows that at an age of 200--250 Myr, Li abundances have already 
converged onto similar values. 
This is true also for M~34 (see Fig.~\ref{M34}): note that, whereas our sample
for NGC~6475 is itself statistically significant for stars 
hotter than $\sim$ 4800~K,
the two samples together allow us to exclude with an even higher
significance that this result is due to low number statistics.
As already evidenced, M~34 is characterized by a Li scatter among stars cooler
than $\sim$ 4700--4800~K: this dispersion could be present also in NGC~6475,
but the NGC~6475 sample is rather
sparse in this temperature range and we cannot draw any definitive conclusion
about this point.
\section{Discussion}
\subsection{The dependence of Li depletion on metallicity}
The issue of the Li--metallicity dependence
has been discussed by several authors
since observational results are in sharp contrast with the predictions of 
both standard and non--standard models.
As mentioned in the introduction,
the models predict that the gas opacity
increases as the iron content increases:
therefore, one expects the convective envelope to
reach more internal layers, and Li destruction to be
more efficient in higher [Fe/H] clusters.
At the same time, the opacity values are also 
affected by the abundance of
oxygen (and other $\rm{\alpha}$ elements): in particular, an enhanced
O/Fe content should
move the base of the CZ towards deeper layers (Piau \& Turck-Chi\`eze
\cite{pt02}).

A clear effect of [Fe/H] on Li depletion has never been empirically
confirmed:
for example, the comparison between the Pleiades ([Fe/H]$\sim$solar)
and \object{Blanco~1} ([Fe/H]$\rm{=+0.14}$, 
Jeffries \& James \cite{blanco1}), both with an age of $\sim$100 Myr,
suggests that PMS Li depletion does not depend on metallicity.

We found
NGC~6475 to have [Fe/H]$\rm{=+0.14\pm0.06}$; as mentioned
in the introduction, the metallicity
of M~34 is somewhat lower, but most likely not as low
as found in the early study of Canterna et al.~(\cite{canterna}):
in fact, 
Schuler et al.~(\cite{schuler}) derived for this cluster
[Fe/H]$\rm{=+0.07\pm0.04}$.
This result is based on five solar--type stars,
while, considering their whole sample of nine stars with
$\rm{{4750}\leq{T_{eff}}\leq{6130}}$ K, Schuler et al. would have derived
[Fe/H]$\rm{=+0.02 \pm 0.02}$; excluding only the
two coolest stars of the total sample, they would have instead found 
[Fe/H]$\rm{=+0.04}$ for M~34.
The value of the iron content of this cluster is a
very crucial point for our
discussion: as seen in Fig.~\ref{M34},
there are no significant differences between the 
\nli~distributions of NGC~6475
and M~34, for stars hotter than $\sim$ 4800~K.
The uncertainty on the metallicity of M~34 leads to two different
possibilities for
the interpretation of this result: 
(i) if M~34 has [Fe/H]$\rm{=+0.07\pm0.04}$ (as probable, given the
more recent and
detailed analysis of Schuler et al.~\cite{schuler}),
the similarity between the Li distribution of this cluster and NGC~6475
would not be surprising;
(ii) if, on the contrary, M~34 has a lower, close to solar, metallicity,
the results of
Jeffries \& James~(\cite{blanco1}) based on the comparison
of Blanco~1 and the Pleiades would be extended to larger ages,
i.e.
the overall metallicity does not affect Li depletion
up to the age of NGC~6475.

In any case,
both possibilities allow us to safely merge the NGC~6475
and M~34 samples to investigate Li evolution as a function of age
by comparing these clusters with the younger
Pleiades and the older Hyades:
in fact, in case (i) we can use the Pleiades in the comparison,
since their Li distribution is similar to
that of the over--solar metallicity Blanco~1.
We will use the Pleiades instead of Blanco~1 since  
a very rich sample is available
for the former cluster, allowing also a more detailed discussion
about the spread among K--type stars; if, otherwise, case (ii)
is the correct one,
the Li patterns of
NGC~6475 and M~34 are not affected by metallicity, thus age is the main
parameter on which MS Li depletion depends.
Finally and obviously, in both cases there is no problem in using the Hyades
([Fe/H]$\rm{=+0.13}$) in our comparison.

The above discussion is valid for stars warmer than $\sim$4800 K;
cooler stars deserve a special remark,
since these stars in M~34 show a scatter 
in Li abundances: as mentioned,
we cannot draw any definitive conclusion
about the presence of a similar spread in NGC~6475.
Under case (i), 
i.e. similar [Fe/H] for the two clusters,
one would expect to find a scatter in NGC~6475.
If M~34 has instead a solar metallicity (case (ii))
and the scatter exists also
in NGC~6475, 
this would mean that the iron content does not
affect Li depletion even for the coolest stars, at least at an age of 
$\sim$220--250 Myr.
On the contrary,
if further Li observations of NGC~6475
should demonstrate that no spread is present in this cluster,
this would suggest that Li depletion in cool stars is affected by metallicity
and the mechanism
causing the dispersion in Li is also metal dependent.

Whereas we leave the issue of the spread among stars cooler than 
$\sim$ 4800~K to a future
larger sample, we discuss below our results for the hotter stars.

\subsection{Early--MS Li depletion}
\subsubsection{The observed time scales of Li depletion}

\begin{figure}
\psfig{figure=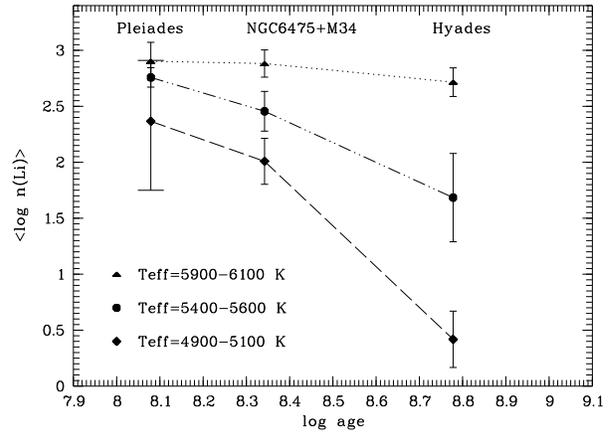, width=6.0cm, angle=-90}
\psfig{figure=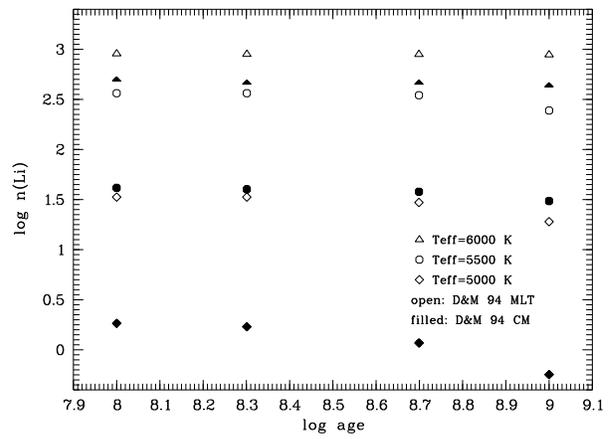, width=6.0cm, angle=-90}
\caption{Li abundance vs.~$\log$ age for three ranges
in \teff~($\mathrm{6000\pm100}$ K, $\mathrm{5500\pm100}$ K,
$\mathrm{5000\pm100}$ K).
In the top panel average \nli~values for the Pleiades, NGC~6475+M~34
and the Hyades are shown. 
Error bars represent $\mathrm{1\sigma}$ standard deviations
from the average value; for the Pleiades
in the range $\mathrm{5000\pm100}$ K, where the Li spread is present,
we have plotted the maximum and minimum values of~\nli.
In the bottom panel theoretical \nli~values 
at various ages are plotted (D'~Antona
\& Mazzitelli 1994), including two different treatments of
overadiabatic convection: MLT is the mixing length theory model, CM
is the Canuto \& Mazzitelli model.}\label{mean}
\end{figure}
The NGC~6475 merged sample confirms that
MS Li depletion occurs for stars cooler than $\sim 6000$~K, both between
120 and 220 Myr and between 220 and 600~Myr.
In order to infer
the time scales of Li depletion for stars of different temperatures
we plot in the upper panel of Fig.~\ref{mean}
the average Li abundance as a function of age,
for the Pleiades, the Hyades and the NGC~6475$+$M~34 merged sample; 
error bars represent $\mathrm{1\sigma}$ standard
deviations from the average values. 
Three different temperature ranges are considered: (i)~\teff$\mathrm{=6000\pm100}$ K,
(ii)~\teff$\mathrm{=5500\pm100}$ K, and (iii)~\teff$\mathrm{=5000\pm100}$ K.
For range (i) Li destruction occurs very slowly
both between 120 and 220 Myr and between 220 and 600 Myr; 
in range (ii), Li depletion appears as a continuous process
and it seems to be linearly related to $\log t$,
meaning that Li abundance scales as n(Li)${=t^{-\alpha}}$;
we found $\rm{\alpha=1.15}$ in the age interval
[120--220] Myr and a slightly
higher slope for the range [220--600] Myr
($\rm{\alpha=1.77}$), suggesting that Li depletion accelerates
after 220 Myr.
The last \teff~range is more difficult to deal with because of
the presence of the Li spread for the Pleiades.
First, note that for the Pleiades (in this \teff~range)
the plotted error bar represents the difference between
the maximum and minimum Li abundances
(instead of the standard deviation).
Second,
if one considers the average Li abundance
for stars with \teff~around 5000~K, Li destruction
appears much more efficient between 220 and 600 Myr than
between the age of the Pleiades and that of NGC~6475. In fact, in this case
the exponents of the depletion law are $\rm{\alpha=1.36}$ 
([120--220] Myr) and $\rm{\alpha=3.65}$ ([220--600] Myr),
meaning that the mixing process
undergoes a large acceleration for the
coolest stars after $\sim$ 220 Myr.
On the other hand, considering the mean \nli~value for Pleiades stars
with \teff~around 5000~K may not be 
very significant, since, given the amount of scatter,
the average is probably not a
representative quantity.
If we look instead at the evolution of Li from the
maximum and minimum values in the Pleiades, up to the age of the Hyades,
two different scenarios are possible:
1) Evolution from the upper envelope of the Pleiades up to the Hyades age:
Li depletion occurs very rapidly
and with nearly constant time scale from 120 Myr to 600 Myr; 
2) Evolution from the lower envelope of the Pleiades 
up to the Hyades age: very little (if any) depletion occurs
between $\sim$ 120 and 220 Myr, followed by
a fast Li depletion
between 220 Myr and the Hyades age.
\subsubsection{Comparison with standard models}\label{comparison}
In order to adequately discuss these observational features, we
compare the observational scenario with quantitative theoretical 
predictions of standard models. 
In the bottom panel of Fig.~\ref{mean}
we plot Li abundances predicted by the standard models of 
D'~Antona \& Mazzitelli (\cite{dm94}) for Pop.~I stars ($\rm{Z=0.019}$,
$\rm{Y=0.028}$)
with ages ranging from 100 Myr to 1 Gyr; the three temperature ranges
are the same as in the top panel of Fig.~\ref{mean}.
Two different models are considered: in one model (MLT) the treatment
of overadiabatic convection relies on the Mixing Length Theory,
while the other (CM) includes the Canuto \& Mazzitelli overadiabatic
convection model (see D'~Antona \& Mazzitelli \cite{dm94}
and references therein for further details).
The depletion patterns are almost flat for
the three temperature ranges, indicating
that little depletion is expected after arrival on the ZAMS.
We mention that,
whereas the absolute amount of PMS depletion 
(and thus the value of \nli~at 100 Myr)
depends on the treatment of
overadiabatic convection, the relative amount of MS
depletion which we are interested in is very small in all cases.

The comparison of the two panels of Fig.~\ref{mean} shows that:
(i) 6000~K-- observations agree with the theoretical
predictions. The models
predict, during both the PMS and MS phases, a
temperature at the base of the convective envelope ($\rm{T_{CZ}}$)
which is slightly lower than (or at most similar to)
the Li burning temperature ($\rm{T_{Li}=2.5\times10^{6}}$ K).
The agreement between observations and model predictions
allows us to conclude that in late F stars
no extra--mixing mechanism is probably present, at least up to the Hyades age;
(ii) 5500~K-- there is a clear disagreement between
theory and observations: the models predict
$\rm{T_{CZ}}\sim3\times10^{6}$ during the PMS,
but $\rm{T_{CZ}}$ decreases down to values
around $\rm{T_{Li}}$
before an age of $\sim$ 100 Myr, thus very little Li depletion
is present after this age\footnote{We mention that, within a model,
the Li burning efficiency strictly depends on the assumed
nuclear reaction rates; thus, even if $\rm{T_{CZ}\sim{T_{Li}}}$,
the Li burning could be poorly efficient.}.
Standard models cannot explain
the observed MS depletion: this confirms that an extra--mixing mechanism
is at work in these stars. 
As suggested by several authors
(see Jones et al.~\cite{jones97} and references therein) extra--mixing could be
due to MS angular momentum loss (AML) and angular momentum
transport, which, in this case should
be a continuous process. If, however,
Li depletion is driven by rotational mixing,
it is difficult to understand the lack of dispersion among these stars
which have, presumably, different rotational histories.
(iii) 5000~K-- During the PMS, the theoretical $\rm{T_{CZ}}$ is higher
than the Li burning temperature, even at very young ages ($\sim$ 0.3 Myr),
but it decreases after arrival on ZAMS; thus,
according to the models, 100 Myr old cluster stars should have depleted
a large amount of Li during the PMS. 

With regard to the latter point
we can consider two opposite hypotheses:
{\it a)} the upper envelope of the
Pleiades is the result of PMS convection only and the lower envelope is
over--depleted by the action of an
extra--mixing mechanism during PMS;
{\it b)} the lower envelope of the Pleiades
is the result of convection only, while
in Li rich Pleiades stars the PMS convection 
might have been strongly inhibited by some non--standard process;
for example, as suggested by 
Ventura et al. (\cite{ventura}) and D'~Antona et al. (\cite{dantona00}),
the effect of magnetic fields induced by a strong rotation
could inhibit Li depletion;
alternatively, a strong rotation could significantly 
modify the stellar structure
(see Mart\'\i n \& Claret \cite{claret}) and prevent
convection and Li depletion:
both the proposed processes 
could explain the dispersion since both rotation and magnetic fields 
cover a large range of values within the same cluster (e.g.
Stauffer et al.~\cite{stauffer}).

Case {\it a)} appears unlikely since
it would imply that, during the PMS, the convection
in these cool stars ``normally'' do not reach deep enough layers
to burn lithium.
Hypothesis {\it b)} appears more probable;
within this hypothesis one can explain both the
convergence of Li abundances at the age of NGC~6475 and the large differences
in Li depletion time scales in the two intervals [100--220] Myr
and [220--600] Myr with the following speculative scenario:
in stars of the upper envelope Li depletion is inhibited during PMS
(by magnetic fields and/or rotation, see above); then, after the stars
have reached the ZAMS ($\sim$ 100 Myr),
they start loosing angular momentum at a fast rate
and extra--mixing
occurs, leading to MS Li destruction which
is a continuous process from 100 Myr to 600 Myr.
Stars on the lower envelope (which are mostly
slow rotators) during the PMS
deplete a large amount of Li under the action of convection only, which stops
at an age of $\sim$ 100 Myr; Li depletion becomes again 
sensitively efficient
around the age of NGC~6475, when the decoupling between the core
and the surface is large enough to have extra--mixing due to AML.
Note that these cool stars have rather deep convective envelopes
and that during MS their $\rm{T_{CZ}}$ remains very close to the temperature
which makes Li burning efficient: therefore only a small amount
of extra--mixing is required for Li depletion to occur and this explains
the large efficiency of Li depletion between 220 and 600 Myr,
for stars of both the upper and lower envelopes.

In summary, we suggest that the spread observed among Pleiades stars
cooler than $\sim$ 5500~K could be due to the fact that
convection and Li depletion may be inhibited by processes related to rotation
and/or magnetic fields, which vary from star to star;
we also conclude that the convergence of Li abundances at
the age of NGC~6475 for stars hotter than $\sim$ 4700--4800~K
could be due instead to extra--mixing mechanisms, which drive the depletion
after arrival on the ZAMS, and have different time scales depending on
the initial rotation.
Finally, we stress that in the discussion above we assumed that
the spread observed in the Pleiades
and M~34 is due to a real scatter in Li abundances.
We mention that several authors
suggested that
the spread in Li equivalent widths could not necessarily
correspond
to a real spread in abundances, and they
investigated whether the scatter could be due to the effects
of surface activity (spots in particular) 
on the line formation and strength
(e.g. Stuik et al.~\cite{stuik}; King et al.~\cite{king}; Randich \cite{potassio}; Barrado y Navascu\'es
et al.~\cite{barrado}).
The issue of the scatter in Li abundances remains therefore open.
\section{Conclusions}
We have obtained high resolution CASPEC spectra for
34 late F to K--type stars in the young open cluster
NGC~6475 (age $\sim$ 220 Myr, intermediate between the Pleiades and 
the Hyades), thus extending
the previous observations of JJ97
and J00. For a large part of the stars
we derived CORAVEL information on membership and binarity:
26 stars turned out to be probable cluster members, while
only one star turned out to be non--member.
The other 7 stars were not observed with CORAVEL, but
given their spectral characteristics and
Li abundances, we considered them as probable members.

Our main results are:\\
{\it a)} We confirm the over--solar metallicity of the cluster; specifically
we found [Fe/H]$\rm{=+0.14\pm0.06}$.\\
{\it b)} The comparison of NGC~6475 with 
the similar age M~34 shows no significant differences
between the two Li distributions down to \teff$\sim$4700--4800 K. This is not
surprising, given the small difference in metallicity between
the two clusters, according to Schuler et al.~(\cite{schuler})
which found [Fe/H]$\rm{=+0.07\pm0.04}$ for M~34.
M~34 shows an evident Li abundance spread
among stars cooler than $\sim$ 4700--4800~K; the NGC~6475 sample
is instead rather sparse in this temperature range.
Thus, although there may be
an indication for the presence of a dispersion, no definitive conclusion
can be drawn. More late K--type stars
in NGC~6475 should be observed.\\
{\it c)} Assuming that metallicity does not affect PMS Li depletion,
as shown by the Pleiades--Blanco~1 comparison,
we can consider the age sequence from the
Pleiades, to NGC~6475+M~34 and then to the Hyades. We found that
Li depletion occurs during the MS phase
of G and K--type stars;
the Li pattern of NGC~6475 for both
G and early K stars lies between
those of the Pleiades and of the Hyades. This means that extra--mixing
mechanisms are likely at work both between $\sim$ 100 and $\sim$ 220 Myr
and between $\sim$ 220 and $\sim$ 600 Myr.\\
{\it d)} The star--to--star scatter in Li abundance observed
among stars cooler than $\sim$ 5500~K in clusters as young as 
(and younger than) the Pleiades
is not present in NGC~6475 stars hotter than $\sim$ 4700--4800~K,
as well as in M~34 stars over the same temperature range.
We suggest that the spread observed in the Pleiades could be due to
processes related to rotation and magnetic fields, which inhibit
convective mixing and Li depletion during the PMS for
part of the stars (those in the upper envelope); 
the disappearing of the scatter 
at the age of NGC~6475 (for stars
in the temperature range [5500--4700] K) could be due to extra--mixing
processes, which could also be responsible for the 
acceleration of Li depletion
between this age and that of the Hyades.

As a final remark, we stress that the determination of
oxygen and other $\rm{\alpha}$ elements abundances
are a very important issue for the investigation of Li evolution.
\begin{acknowledgements}
This work has partially been supported 
through a grant by Ministero dell'Istruzione, Universit\`a e Ricerca (MIUR)
to S.~Randich and R.~Pallavicini.
We thank the referee, Dr. B. F. Jones, for his very useful comments.
\end{acknowledgements}
{}

\end{document}